\documentstyle[11pt,newpasp,twoside,epsf]{article}
\def\edcomment#1{\iffalse\marginpar{\raggedright\sl#1\/}\else\relax\fi}
\reversemarginpar

\begin{document}
\title{The low mas end of the young cluster IC2391}
 \author{David Barrado y Navascu\'es}
\affil{LAEFF-INTA. Apdo. 50727, 28080 Madrid. SPAIN}
\author{John R. Stauffer}
\affil{IPAC, California Institute of Technology,
         Pasadena, CA 91125, USA}


\begin{abstract}
By collecting optical and infrared photometry, as well
as medium resolution spectroscopy, we have discovered a 
sample of low mass stars and brwon dwarfs in the young
cluster IC2391. Using the lithium depletion boundary
near the substellar limit, we have estimated the age
of the cluster as $\sim$50 Myr. We have also estudied
the H$\alpha$ emission in this sample.
\end{abstract}


\section{Introduction}

During the last few years, a large amount of knowledge has been
gained regarding the nature and  properties of low mass stars and   
brown dwarfs. We have studied several nearby, young clusters,
 which, due to these characteristics, are excellent 
targets. One of these clusters is IC2391, 
located at 155 pc
--(m-M)$_O$=5.95$\pm$0.1--, and with a low interstellar reddening
--E($B-V$)=0.06 (Patten \& Simon 1996).
In previous papers, we have estimated the age
of the cluster based on the lithium depletion technique
($\tau$=53$\pm$5 Myr, Barrado y Navascu\'es et al. 1999)
and identified  a large number of low mass candidate 
members (Barrado y Navascu\'es et al. 2001).
In this paper, we present new spectroscopy 
of cluster members, improve the lithium  age, 
and analyze other properties such as the H$\alpha$
activity and the mass function (MF).

\section{Analysis}

\subsection{The new spectroscopic data.}

We have collected multifiber medium resolution spectroscopy with the
Hydra II spectrograph at the CTIO 4m telescope in March 10th  and 13th, 1999.
We took several individual exposure of 1 hour each (except the last one, of 
half an hour), totaling 11.5 hours. Each of these exposures
was processed one by one, and the individual spectrum extracted 
within the IRAF environment. Finally, they were combined 
to provide a single spectrum for each  target.
For each IC2391 candidate member, the 1 hour spectra and the combined 
spectrum were used to derive rough radial velocity, H$\alpha$
and lithium equivalent widths --W(H$\alpha$) and W(LiI6708 \AA).
We also estimated the spectral types
based on comparison with field stars and IC2391 members with
previously known classification.

\begin{figure}
\plotfiddle{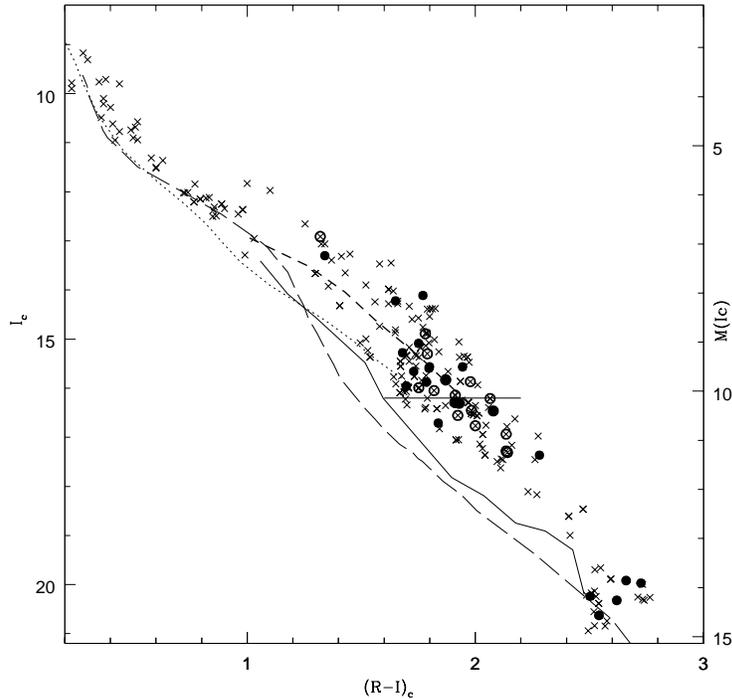}{8.0cm}{0}{50}{50}{-150}{-95}
\caption{Color-magnitude diagram for IC2391 candidate members.
Crosses represent all the available photometric data from 
Simon \& Patten (1998), Patten \& Pavlovsky (1999) 
and Barrado y Navascu\'es et al. (2001).
Open circles correspond to spectroscopic data from 
Barrado y Navascu\'es et al. (1999), whereas solid circles
were observed with HydraII. Several 50 Myr isochrones (see
text) and an empirical zero age main sequence are
also represented. }
\end{figure}


Figure 1 displays a color-magnitude diagram for the cluster.
All  known IC2391 photometric candidate members are shown as
crosses, whereas  circles represent objects which have spectroscopy
(open and solid from Barrado y Navascu\'es et al. 1999 and 
this paper, respectively). 

The previously identified location of the lithium
depletion boundary (LDB) is indicated with a solid horizontal segment.
The new data, combined with the former 
spectroscopic, allow a detailed study of the 
depletion of lithium near the substellar limit and, hence, to improve the 
age estimate of the cluster based on this technique. In addition, 
we have collected spectra of several low mass brown dwarf candidates,
 with M($Ic$)$\sim$14 mag, establishing membership in about half of them.
Jameson et al. (2002) has notice that there is a 
gap in the CMD in several young clusters for mid-M spectral type members.
Our IC2391 data appear to support its reality.

\subsection{Lithium depletion and the age of the cluster}

Figure 2 represents the lithium equivalent widths versus the absolute
$I_C$ magnitude and the ($R-I$)$_C$ color. For completeness, we have
included data from the literature as squares (open symbols
denote upper limits. Our IC2391 data (this paper and 
Barrado y Navascu\'es et al. 1999) are displayed as solid circles and 
open triangles (upper limits).
These two panels clearly show the presence of the so-called  
``lithium chasm'', where lithium has been completely 
depleted,  and the location of a sharp increase in the
W(Li) at the right-hand side, corresponding to the LDB
(dotted line). Note that we have detected lithium in 
two low mass stars located to the left of the LDB.
A similar situation has been found in the Pleiades
cluster (Oppenheimer et al. 1997).  For our two stars in
IC2391, the fainter of the two is most easily explained as
a nearly equal mass binary; the brighter star with M(I$_c$) $\sim$
8.0 cannot be explained in that fashion (and thus deserves further
study).
The magnitude of the LDB can be estimated as
M($I_C$)=10.15 mag.

\begin{figure}
\plotfiddle{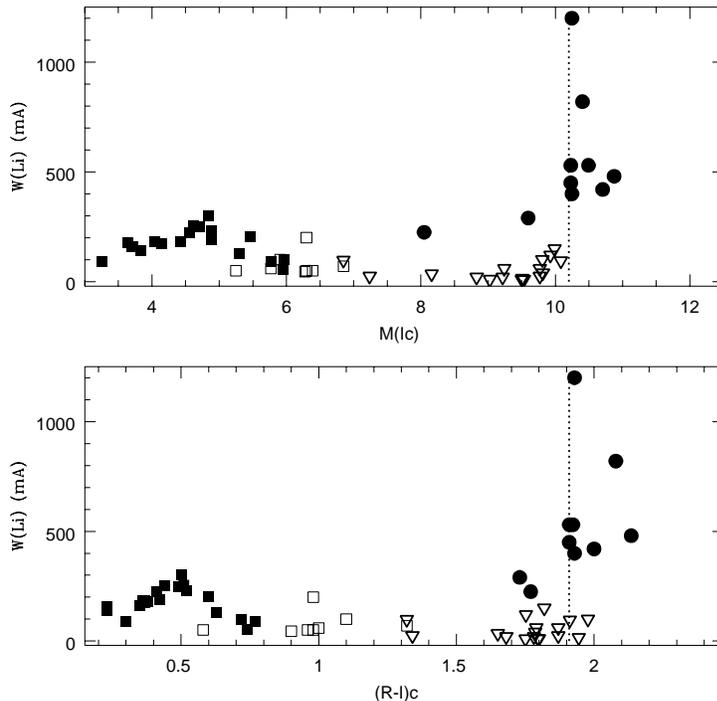}{8.0cm}{0}{50}{50}{-150}{-95}
\caption{Lithium equivalent width versus the absolute $Ic$ 
magnitude ({\bf a}) and $(R-I)c$ color index  ({\bf b}).
Circles and triangles represent data from this work and 
Barrado y Navascu\'es et al. 1999, whereas squares correspond to data from
the literature. Actual data and upper limits  are displayed as solid  
and  open symbols, respectively.
The vertical dotted line locate the lithium depletion boundary
for the cluster.
 }
\end{figure}

Another color-magnitude diagram is presented in Figure 3.
In this case, we have plotted the 2MASS $Ks$ magnitudes 
against the ($I_C-Ks$) color index. 
Crosses correspond to IC2391 candidate members for which we
have not obtained spectra. Open circles 
represent probable members with no lithium in their spectrum, whereas
solid circles are  probable members where we have been able to identify it. 
We have included a 50 Myr isochrone by Baraffe et al. (1998)  and a
ZAMS (long-dashed and solid lines, respectively).
As happens in Figure 2, the LDB is clearly seen, in this case located at
M($Ks$)=7.596 mag.

We have estimated lithium abundances for the available data.
Figure 4 shows these abundances against the effective temperature.
We have used the Soderblom et al. (1993) curves 
of growth for the stars located at the left-hand side of the
 ``lithium chasm'' and those published by Zapatero
Osorio et al. (2002) for the cool end of the cluster sequence.
Effective temperatures were derived based on Bessell (1979) 
for stars warmer than M0 and 
Leggett (1992) for cooler
members.

Several lithium depletion isochrones
by D'Antona \& Mazzitelli (1994) and Chabrier
et al. (2000) are also included. Upper limits are
shown as solid triangles, whereas solid circles
represent actual abundances.
Although these values are not very accurate for 
IC2391 members close to the substellar regime (due
to problems such as opacities with the atmospheric models),
the LDB is, again, easily visible, at about T$_{eff}$=3000 K.
The isochrone which better reproduce the lithium depletion 
pattern in the IC2391 cluster is a 50 Myr model.

\begin{figure}
\plotfiddle{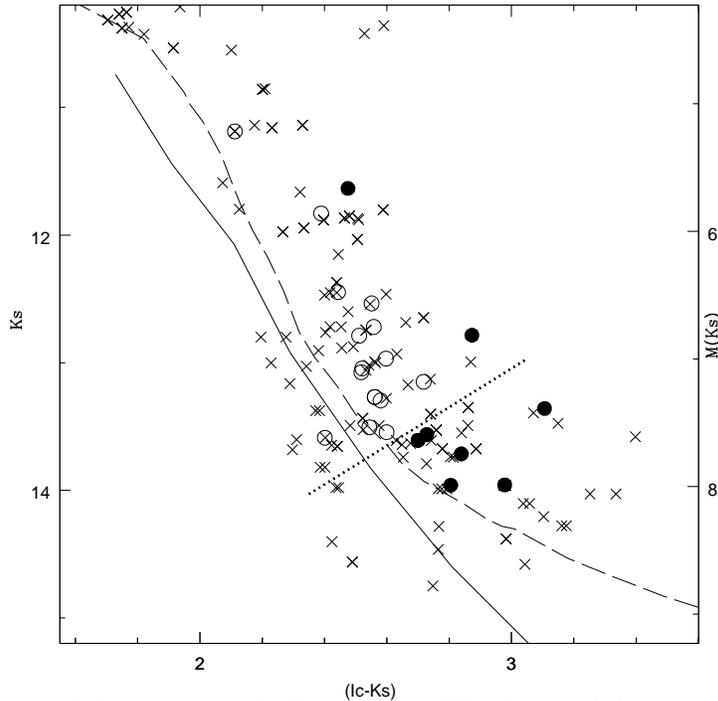}{8.0cm}{0}{50}{50}{-150}{-95}
\caption{Color-magnitude diagram for IC2391 candidate members.
Crosses represent all candidate members from 
Simon \& Patten (1998), Patten \& Pavlovsky (1999) 
and Barrado y Navascu\'es et al. (2001).
IC2391 members with lithium detection are
shows as solid circles, whereas those lacking the Li6708 \AA { }
feature are displayed as open symbols.
A 50 Myr isochrones is also included (Baraffe et al. 1998, long-dashed line),
 as well as an empirical ZAMS (solid line) and the location of the
lithium depletion boundary (dotted line).
}
\end{figure}

So far, we have identified the location of the 
LDB in three open clusters, namely IC2391
(Barrado y Navascu\'es et al. 1999 and this paper),
 $\alpha$ Per (Stauffer et al. 1999)
and the Pleiades (Stauffer et al. 1998).
The standard ages of these clusters, based on
upper main sequence isochrone fitting are $\sim$30, $\sim$50
and 80--100 Myr, respectively.
We have re-analyzed the optical and infrared available data
and establish the LBD in the M($I_C$), M($K$) and  M(bol)
magnitudes. 
Figure 5 displays the comparison for these three magnitudes and 
the three clusters. The solid thick line correspond to 
Baraffe et al. (1998) theoretical tracks, although similar results
can be achieved using Burrows et al. (1997) or D'Antona \& Mazzitelli
(1994) calculations.
The diagram easily shows that, indeed, the lithium ages
are about 50 \% older than the ages derived using the location in 
CMD of  massive stars evolving off the main sequence.
The age difference may be attributable to 
a moderate core-overshooting in the massive stars
(Ventura et al. 1998), which 
would increase their life inside the main sequence 
by a 50\%, though we cannot be certain yet if this is the
correct explanation.

\begin{figure}
\plotfiddle{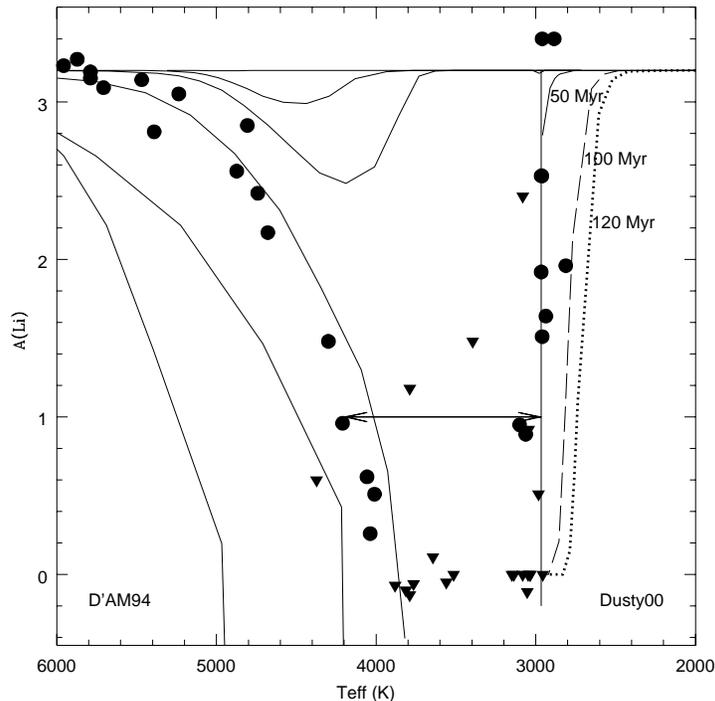}{8.0cm}{0}{50}{50}{-150}{-95}
\caption{Lithium abundance versus effective temperature.
Actual  abundances and upper limits are shown as circles
and triangles, respectively. Several lithium depletion  isochrones
from D'Antona \& Mazzitelli (1994) --1, 3, 5, 10, 20 and 100 Myr, left-- and 
Chabrier et al. (2000) --50, 100 and 120 Myr; right-- are included.}
\end{figure}

\begin{figure}
\plotfiddle{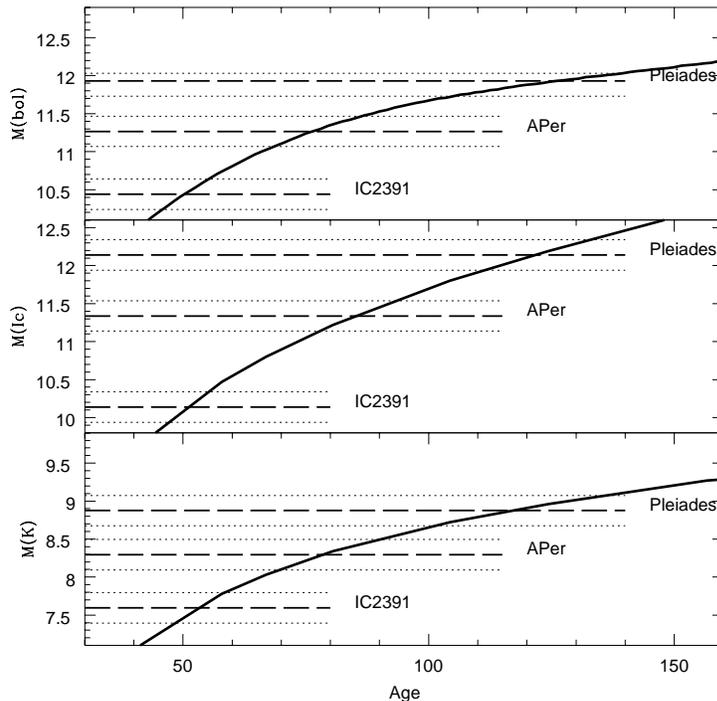}{8.0cm}{0}{50}{50}{-150}{-95}
\caption{Location of the LBD for IC2391, Alpha Per and the Pleiades.}
\end{figure}

\subsection{The H$\alpha$ emission}

We have investigated the H$\alpha$ emission at the end of the 
cluster sequence and compare it with other young clusters.
Figure 5 illustrates the W(H$\alpha$) against the 
($R-I$)$_C$ color index.
IC2391 data are shown as solid circles, whereas 
$\sigma$ Orionis and $\alpha$ Persei data are 
displayed as solid triangles and
crosses, respectively (see Barrado y Navascu\'es et al.
2002 for additional information).
Note that several $\sigma$ Orionis low mass stars and
BDs have very large W(H$\alpha$), located outside the 
diagram, and are not shown for clarity. Actually, 
some of them also have infrared excesses and/or
forbidden lines and have been classified 
as classical TTauri substellar analogs
(Barrado y Navascu\'es et al. 2002).
The comparison between these three clusters
indicates that they have similar pattern 
in the H$\alpha$ emission. In principal, for most of this
objects, the origin of this phenomenology should be
chromospheric, except in the few cases already 
pointed out in $\sigma$ Orionis, which is much 
younger than IC2391. In this last case, as in $\alpha$ Per, 
all the  possible circumstellar disks have had enough time to
dissipate. 

Figure 6  shows two main features:
the steadily increase of the W(H$\alpha$) toward
cooler objects, up to ($R-I)_C$$\sim$1.5 (mid M spectral type),
and a large dispersion for cooler objects, well 
within the substellar domain. Note, however, that the signal-to-noise
ratio is worse for those objects, due to their faintness, and therefore, 
the errors in the W(H$\alpha$) are larger.
The enhanced spread in W(H$\alpha$) for the cooler objects may
also be indicative of larger or more frequent flares for these
objects.

\begin{figure}
\plotfiddle{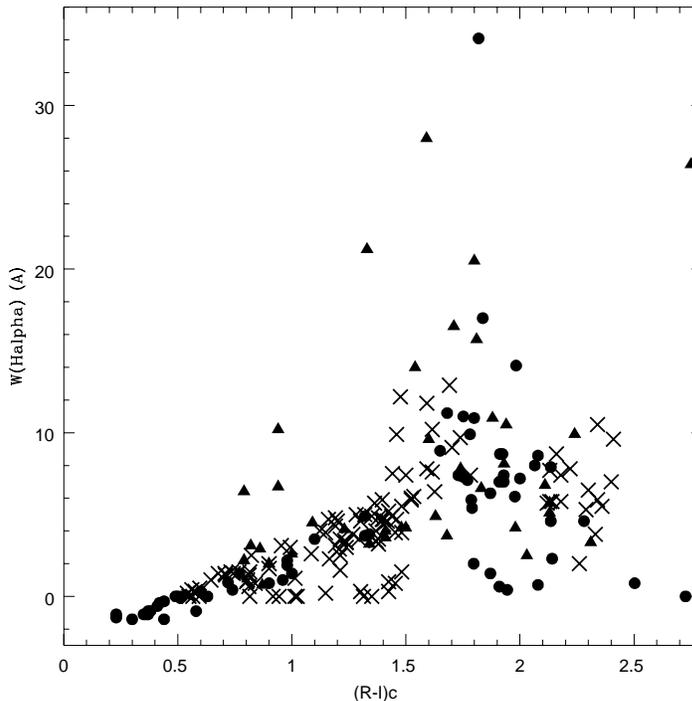}{8.0cm}{0}{50}{50}{-150}{-95}
\caption{Comparison between the H$\alpha$ equivalent widths versus the 
$(R-I)c$ color index for several clusters. Triangles, circles, and crosses
represent data from Sigma Orionis, IC2391 and Alpha Persei clusters.
}
\end{figure}

\subsection{The Mass Function}

We have derived the mass function  of the IC2391
cluster and compared it with other clusters with 
ages ranging from 3 Myr up to 200 Myr
 ($\sigma$ Ori, $\alpha$ Per, the Pleiades, NGC2516 and M35).
Figure 7 contains these mass functions together 
the spectral index of a power law fit for each of them.
All these MFs were derived with Baraffe et al. (1998) 
isochrones. Note, however, that in all cases except for
IC2391 and NGC2516, we were able to remove the
pollution by interlopers using additional data
(infrared photometry and optical spectroscopy). Therefore, 
the index derived for these two clusters are upper limits.
In any case, although there are some differences in the
structure of the MF, the general shape in these clusters
is similar. The only exception is M35, a very rich cluster, 
where some dynamical evolution, with mass segregation, 
might have taken place.

\begin{figure}
\plotfiddle{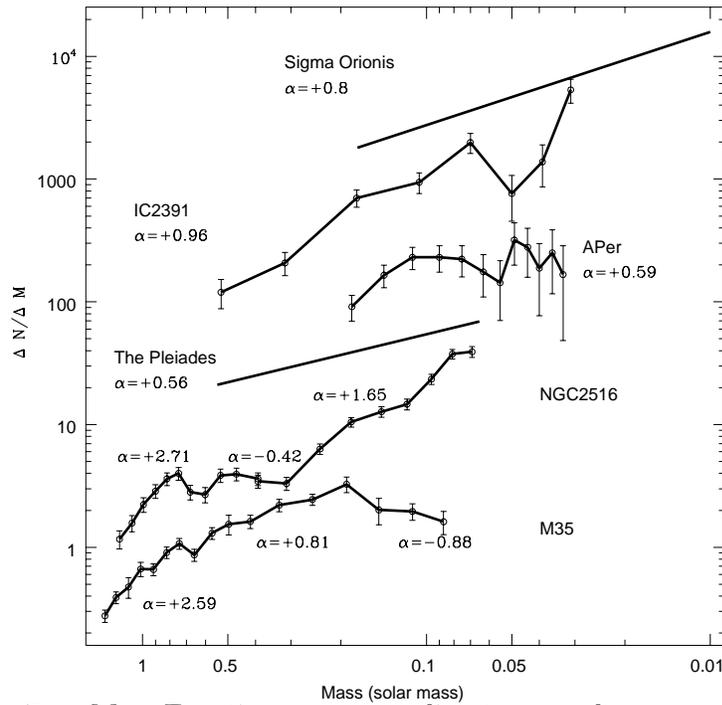}{8.0cm}{0}{50}{50}{-150}{-95}
\caption{Mass Functions corresponding to several 
young open clusters.}
\end{figure}

\section{Conclusions.}

We have derived membership for a sample of IC2391 
low mass stars and BD candidate members using medium
resolution spectroscopy. In addition, by combining these
data with previous spectroscopy, we have re-evaluated the 
LDB of the cluster and its age, which is about 50 Myr.
The distribution of the W(H$\alpha$) is very similar
to those characteristics of younger and older clusters,
as well as the cluster MF.



\end{document}